\documentclass[letterpaper, twocolumn, final]{IEEEtran}

\usepackage{amssymb,amsmath, graphicx}

\DeclareGraphicsExtensions{.pdf,.eps,.ps,.png,.jpg,.jpeg}

\usepackage {bm} \usepackage {tabulary}
\begin{document}
\title{Determining the Number of Samples Required to Estimate Entropy in Natural Sequences
}
\author{Andrew D. Back, \textit{Member, IEEE}, Daniel Angus, \textit{Member, IEEE} and

Janet Wiles, \textit{Member, IEEE}
\\
School of ITEE, The University of Queensland, Brisbane, QLD, 4072 Australia.
}
\maketitle
\begin{abstract}Calculating the Shannon entropy for symbolic sequences has been widely considered in
many fields. For descriptive statistical problems such as estimating the N-gram entropy of English
language text, a common approach is to use as much data as possible to obtain progressively more
accurate estimates. However in some instances, only short sequences may be available. This gives
rise to the question of how many samples are needed to compute entropy. In this paper, we examine
this problem and propose a method for estimating the number of samples required to
compute Shannon entropy for a set of ranked symbolic ``natural'' events. The
result is developed using a modified Zipf-Mandelbrot law and the
Dvoretzky-Kiefer-Wolfowitz inequality, and we propose an algorithm which
yields an estimate for the minimum number of samples required to
obtain an estimate of entropy with a given confidence level and degree of accuracy.
\end{abstract}

\section{Introduction
}
Consider a sequence of symbolic information given by
$\Upsilon _{i} =\left [\upsilon _{0} ,\ldots  ,\upsilon _{n -1}\right ]^{T} ,$
where
$\upsilon _{j} \in \Sigma ^{ \ast } ,$
and
$\Sigma ^{ \ast }$
is an alphabet or finite nonempty set with symbolic members. Suppose we are interested in
the information content in such a message sequence. One way to approach this problem is by measuring
what is new or novel in a given sequence. If a string consists of symbols
$aaa ,bbb ,aaa$
- beyond the first few words, there is little novelty or `surprise' about the message. On
the other hand, if a string consists of symbols
$acb ,bde ,zxy ,eqa$
- then it is evident that the message has a higher degree of novelty. The idea that the
randomness of a message can give a measure of the information it conveys formed the basis
of Shannon's entropy\protect\footnote{For convenience we will generally refer to Shannon Entropy as simply entropy
with the specific formulation evident from the context.
} theory which gives a means of assigning a value to the information
carried within a message \cite{Shannon48},\cite{Shannon483}. The way in which Shannon formulated this principle is that, given a single
random variable
$x$
which may take
$M$
distinct values, and is in this sense symbolic, where each value occurs independently with
probability
$p\left (x_{i}\right ) ,$
$i \in [1 ,M] ,$
then the single symbol Shannon entropy is defined as:
\begin{equation}H_{1}(X) = -\sum \limits _{i =1}^{M}p(x_{i})\log _{2}(p\left (x_{i}\right ))
\end{equation}

This extends to the case where the probabilities of multiple symbols
occurring together are taken into account. Hence,
$H_{2}$
indicates the entropy from the probabilities of two symbols occurring consecutively. The
general \textit{N}-gram entropy, which is a measure of the information due to the
statistical probability of
$N$
adjacent symbols occuring consecutively, can be derived as
\begin{equation}H_{N}(X\vert B) = -\sum \limits _{i ,j}^{_{}}p(b_{i} ,x_{j})\log _{2}(p(x_{j}\vert b_{i}))
\end{equation}
where
$b_{i} \in \sum ^{N -1}$
is a block of
$N -1$
symbols,
$x_{j}$
is an arbitrary symbol following
$b_{i} ,$
$p(b_{i} ,x_{j})$
is the probability of the \textit{N}-gram
$(b_{i} ,x_{j})$
,
$p(x_{j}\vert b_{i})$
is the conditional probability of
$x_{j}$
occurring after
$b_{i}$
and is given by
$p(b_{i} ,x_{j})/p(b_{i})$
.

Shannon demonstrated the concept of entropy by applying it to English
text, obtaining estimates of entropy by using a list of 1027
words which were sampled from 100,000 words of English text \cite{Shannon51}. Entropy has since been
applied to a diverse range of problems including word entropy estimation \cite{Barnard55}, statistical keyword detection \cite{Herrera2008}, phylogenetic diversity measurement \cite{Allen2009}, population
biology \cite{Rao1982}, language assessment of Pictish
symbols \cite{LeePic2010}, facial recognition \cite{Liris2011}, and interpreting gene expression data in functional genomics for drug
discovery \cite{Fuhrman:2000:ASE}.

One of the limitations of computing entropy accurately is the dependence on large
amounts of data, even more so when computing \textit{N}-gram entropy. As a concrete
example, in language analytics, estimates of
entropy based on letter, word and \textit{N}-gram statistics have often relied on large data
sets
\cite{Ebeling94}, \cite{Moreno2015}. The reliance on long
data sequences to estimate the probability distributions used to calculate entropy and attempts to
overcome this in coding schemes is discussed in \cite{schurmann-grassberger-96} where they provide an estimate of letter entropy
extrapolated for infinite text lengths.

Various approaches to estimating entropy over finite sample sizes have
been considered. A method of computing the entropy of dynamical systems which corrects for
statistical fluctuations of the sample data over finite sample sizes has been proposed in \cite{Grassberger1988}. Estimation
techniques using small datasets have been
proposed in \cite{Bonachela2008}, and a novel approach for calculating entropy using the idea
of estimating probabilities from a quadratic function of the
inverse number of symbol coincidences was proposed in \cite{Montalvao2012}. An online approach for
estimating
entropy in limited resource environments was proposed in \cite{Paavola2011}. Entropy estimation over
short symbolic sequences was considered in the context of dynamical time series models based on
logistic maps and correlated Markov chains, where an effective shortened sequence length was
proposed which accounted for the correlation effect \cite{lesne2009}.

A question which naturally arises then is how many samples are required in order to
obtain an accurate estimate of entropy according to some criteria? Answering this question may
provide insight into problems where limited data is available and also for online analytical
information theoretic models which seek to limit data, rather than a longer term descriptive
statistical approaches. In this paper, a method is proposed for estimating the number of samples
required to calculate entropy of a natural sequence. The proposed model is applied to some example
cases, and the implications of this new approach and potential future work is discussed.

\section{Estimating Samples required for Entropy
}

\subsection{Shannon Entropy           Algorithm
}
The Shannon entropy of a discrete random variable
$X$
with discrete probability distribution
$p_{s}(x ,t)$
is defined as:
\begin{equation}H_{s}(X ,t) = -\sum _{x_{t} \in \mathbf{S}_{t}}p_{s}(x ,t)\log _{2}\left (p_{s}(x ,t)\right )
\end{equation}
where
$H_{s}(X ,t)$
is computed on
$p_{s}(x)$
for
$\overline{x}_{p}(t) =[x(t_{0}) ,\ldots  ,x(t_{p -1})]^{T}$
over
$p$
consecutive samples\protect\footnote{ As a notational convenience, we designate
$t_{p} =t -p$
, where
$x(t -1)$
indicates the value of a variable
$x$
one sample before
$x(t)$
which in some cases is indicated as
$x_{t} .$
Also note that while Shannon entropy is commonly specified using base 2, it is also
possible to formulate entropy using base
$e$
or base 10.
} where
$\mathbf{S}_{t} \subset \mathbf{X} .$
The usual approach to calculating entropy is by estimating
$p_{s}(x)$
. Assuming some theoretical, \textit{true}
values for the probabilities
$p_{s}(x ,t) ,$
the accuracy of
$H_{s}(X ,t) ,$
is determined by the accuracy with which
$\left \{p_{s}(x)\right \}$
is estimated.

For small values of
$p_{s}(x ,t)$
,
$\left \vert \log _{2}p_{s}(x)\right \vert $
becomes large and hence small
$p_{s}(x ,t)$
may contribute significantly to
$H_{s}(X ,t)$
. Now, given a finite set of samples, the accuracy with which
$H_{s}(X ,t)$
can be computed will depend on the accuracy with which
$p_{0}(x)$
can be computed, where the probability of the most infrequent
event occurrence is defined\protect\footnote{We expect there to be a greatest lower bound on the set of probabilities,
but a precise minimum does not necessarily exist.
} as
\begin{equation}p_{0}(x) =\inf \left \{p(x ,r)\vert x \in X ,r \in [1 ,M]\right \} \label{p0inf}
\end{equation}
where the
$M$
probabilities (the alphabet size) are computed from
$N$
observations and the empirical probability is defined as
\begin{equation}N_{0} =\frac{n}{p_{0}(x)_{}} \label{mainN}
\end{equation}

This raises questions of how large
$N_{0}$
should be and is there a relationship between
$M$
and the number of samples
$N_{0}$
required to obtain a specified degree of accuracy with some
level of confidence? Intuitively, one would expect that the larger the alphabet size, then the
greater the number of observations required. In the next sections, we develop a method for
determining the number of samples required to estimate the entropy of natural sequences derived from
a given alphabet.

\subsection{Dvoretzky-Kiefer-Wolfowitz Inequality
}
Given a finite set of randomly sampled iid (independent and identically
distributed) observations
$X_{1} ,\ldots  ,X_{n}$
for which there exists an unknown true distribution function
$F(\lambda )$
where\protect\footnote{ Following historical convention, we will at times use notation where
$P(x)$
refers to probabilities, typically distribution probabilities associated with an
event, and
$p_{i}(x)$
refers also to probabilities, where
$p_{i}(x) =P(X =x_{i}).$
}
\begin{equation}F(\lambda ) =P\left \{X_{j} \leqslant \lambda \right \}
\end{equation}
and an empirical distribution function\protect\footnote{In this section we introduce the notation
$\widehat{p}(x)$
to represent empirical probability.
} is available, defined by
\begin{equation}\widehat{F}_{n}(x) =\frac{1}{n}\sum \limits _{j =1}^{n} \mbox{\boldmath $1$} \negthinspace \{X_{j} \leqslant x\}
\end{equation}
where
$ \mbox{\boldmath $1$}
\negthinspace \{X_{j} \leqslant x\}$
is the indicator function defined as
\begin{equation} \mbox{\boldmath $1$}
\negthinspace \{X_{j} \leqslant x\} =\left \{\begin{array}{cc}1 & \text{if}\thinspace X_{j} \leqslant x \\
0 & \text{otherwise}\end{array}\right .
\end{equation}
The Dvoretzky-Kiefer-Wolfowitz (DKW) inequality \cite{dwk56} extends earlier asymptotic
results by Kolmogorov and Smirnov \cite{doob1949} and provides a probabilistic bound on the difference between the
empirical and true distributions. A tighter probabilistic bound was
obtained by Masaart \cite{massart1990}, which allows the DKW inequality to be expressed as
\begin{equation}P\left \{\sup _{x \in \mathbb{R}}\left \vert \widehat{F}_{n}(x) -F(x)\right \vert  >\epsilon \right \} \leq 2e^{ -2n\epsilon ^{2}}
\end{equation}
Hence, using this inequality, for every
$\epsilon  >0$
, and
$(1 -\zeta ) >0 ,$
it is possible to calculate
$N(\epsilon  ,\zeta )$
such that if
$n \geq N(\epsilon  ,\zeta )$,
\cite{Zielinski2007},
then
\begin{equation}P\left \{\sup _{x \in \mathbb{R}}\left \vert \widehat{F}_{n}(x) -F(x)\right \vert  >\epsilon \right \} \leq 1 -\zeta  .
\end{equation}
A novel application of the DKW inequality to determining a probabilistic upper bound on
the entropy of an unknown distribution based on a sample from that distribution was given recently
by Learned-Miller \& DeStefano \cite{Learned2008}. In the work presented here, we consider the number of samples required to
obtain a specified degree of accuracy with a given confidence level.         Hence, suppose we wish
to determine
$n$
, such that with some degree of confidence
$\zeta$,
the maximum difference between
$\widehat{F}_{n}(x)$
and
$F(x)$
is
$\epsilon$,
then it follows that a solution many be found as
\begin{equation}P\left \{\sup _{x \in \mathbb{R}}\left \vert \widehat{F}_{n}(x) -F(x)\right \vert  <\epsilon \right \} \leq \zeta  .
\end{equation}
Now, the DKW inequality specified a bound for the difference in
distribution functions
$\widehat{F}_{n}(x)$
and
$F(x).$
For discrete random variables with discrete probabilities, we have
\begin{align}\widehat{F}_{n}(r) &  =P(X \leqslant r) \\
 &  =\sum \limits _{k =0}^{r}\widehat{p}(k)\end{align}
Hence, we obtain
\begin{equation}P\left \{\sup _{x \in \mathbb{R}}\left \vert \sum \limits _{k =0}^{r}\left (\widehat{p}(k) -p(k)\right )\right \vert  <\epsilon _{0}\right \} \leq \zeta 
\end{equation}
where,
\begin{equation}1 -\zeta  =2e^{ -2n\epsilon _{0}^{2}}
\end{equation}
and rearranging, this becomes
\begin{equation}n =\frac{1}{2\epsilon _{0}^{2}}\ln \genfrac{(}{)}{}{}{2}{1 -\zeta } . \label{numberN}
\end{equation}
Now, the confidence level is specified by
$\zeta$,
and through consideration of
$H_{s}(X ,t)$
it is clear that
$\epsilon _{0}$
can be specified by the number of samples required to discriminate between the two
closest probabilities used in the entropy calculation through the DKW inequality as
\begin{equation}P\left \{\sup _{x_{r} \in \mathbf{X}}\left \vert \overset{}{\widehat{p}}_{n}(r -1) -\widehat{p}(r)\right \vert  <\epsilon _{r}^{}\right \} \leq \zeta 
\end{equation}
Hence, given a finite set of randomly sampled iid observations
$X_{1},\ldots ,X_{n}$
for which it is assumed that there exists a corresponding set of monotonic probabilities
for a set of
$M$
possible events, i.e.
$\widehat{p}(x_{1}) ,\ldots  ,\widehat{p}(x_{M})$
, then define
\begin{equation}\Delta _{0} =\inf _{i}\left \{\left \vert \widehat{p}(x_{i}) -\widehat{p}(x_{i +1})\right \vert \right \}\qquad i =1 ,\ldots  ,M -1 \label{deltaeqn}
\end{equation}
Now, it follows that
$\epsilon _{r} <\Delta _{0}$,
moreover, it can be observed that if
$\epsilon _{r}^{} =\Delta _{0}/2$,
then at worst, we cannot still reliably discriminate between
$\overset{}{p}_{s}(r)$
and
$\overset{}{p}_{s}(r -1),$
since for this case,
$\sup \left \{\widehat{p}\overset{}{_{c}}(r)_{}\right \} =\inf \left \{\overset{}{\widehat{p}_{c}}(r -1)\right \} ,$
where
\begin{equation}\overset{}{\widehat{p}_{c}}(r) \in \mathbb{R} :\overset{}{\widehat{p}}(r) -\epsilon _{r}^{} \leq \overset{}{\widehat{p}_{c}}(r) \leq \widehat{p}(r) +\epsilon _{r}^{}
\end{equation}
Hence, consider a rule for determining the entropy probabilities as a function of rank,
alphabet size and some parametrization, i.e.
$\overset{}{\widehat{p}}(r) =f(r ,M;\theta ).$
From (\ref{p0inf}), then for a number of observations
$N_{0}$
and expected number of events
$n_{0},$
with the empirical probability defined in (\ref{mainN}), an expression is required for
\begin{align}\epsilon _{r} &  =f\left (\widehat{p}(r)\right ) ,\quad r \in [1 ,M] \\
 &  =f\left (r ,M\right )\end{align}
which will be considered in the next section.

\subsection{Probabilistic Event Model
}
For various natural sequences, the
probability of information events can generally be ranked into monotonically
decreasing order. This phenomena has been examined extensively, in
particular, it was demonstrated by Zipf's early work that the
frequency of ranked words in a text occur in such a way that they can be described by a power law
\cite{Zipf1935}. For natural
language, it has been shown that Zipf's law approximates the distribution of probabilities of letter
or words
across a corpus of sufficient
size for the larger probabilities \cite{Piantadosi2014}. The universality of
Zipf's law has been challenged and in particular, it has been shown to arise as a result of the
choice of
rank as an independent variable \cite{li92random},\cite{liw02}. Nevertheless,
Zipfian laws have been proven to be useful as a means of
statistically characterizing the observed behaviour of symbolic sequences of data \cite{Montemurro2001}.

For the purpose of the development here, we
do not rely on Zipf's law to provide a universal model of human language or other natural sequences
(see for example, the discussions in \cite{liw02},\cite{Corral2015}). Instead, it provides a convenient statistical model which
enables the transformation between the
ranking of symbolic events and an estimate of their expected
probabilities. Hence, it is useful in
forming a model of symbolic information transmission which is
organized on the basis of sentences
made by words in interaction with each other (this may be
considered in a general sense of natural
sequences, not just human language) \cite{Cancho01}. Thus, Zipfian
based models can be useful as a means of viewing the
probabilistic rankings of the symbols employed in natural sequences.

For the calculation of
entropy, the accuracy will depend on the accuracy of calculating the set of probabilities. It
follows that we might expect the accuracy of the probability calculations will be determined by the
most infrequent event occurrences. Therefore, the number of samples required to estimate the
probability of the least frequent event
$p_{0}(r)$
determines the number of samples required to estimate the entropy for the corresponding
set of data.

We proceed by imposing a probabilistic model of the symbolic events for a given
sequence. Since we are dealing with natural sequences of symbolic data, we consider a Zipfian model
approach. The most basic form of Zipf's law models the frequency rank
$r$
of a word\protect\footnote{ A word or N-gram is not necessarily referring to human language, but
indicates a specific set of sequentially occurring symbols.
}, i.e. the \textit{r}-th most frequent word, by a simple inverse
power law, such that the frequency of a word
$f(r)$
scales according to an equation which is approximately
\begin{equation}f(r) \propto \frac{1_{}}{r^{\alpha }}
\end{equation}
where a constant of proportionality dependent on the particular corpus may be introduced,
\cite{li92random} and where
typically
$\alpha  \approx 1.$
Thus, if
$p_{i}(x)$
follows a Zipfian law, then
$p_{0}(x) \propto 1/M$
and
$p_{i}(x) =\varphi f(r)$
. Numerous other variations of this general law have been proposed to provide more
accurate representations, including the Bradford Law, Lotka Law \cite{ChenLeim86}, \cite{ChenLeim90}. A more well known approach
is the Zipf-Mandelbrot law \cite{BOOTH1967}. Given symbols
$x \in \Sigma ^{M +1}_{}$
from an alphabet of size
$M +1$
which includes a blank space
$w_{s}$
, then for any random word of length
$L ,$
given by
$v_{k}(L) =\{w_{s ,}x_{1 ,\ldots  ,}x_{L ,}ws\}$
,
$k =1 ,\ldots  ,M^{L}$
the frequency of occurrence is determined as
\begin{equation}p_{i}\left (L\right ) =\frac{\lambda }{\left (M +1\right )^{L +2}}\qquad i =1 ,\ldots  ,M^{L}
\end{equation}
then Li showed that
$\lambda $
can be determined via the summation of all probabilities of such words \cite{li92random}, hence
\begin{align}\sum \limits _{L =1}^{\infty }M^{L}p_{i}\left (L\right ) &  =1 \\
\sum \limits _{L =1}^{\infty }M^{L}\frac{\lambda }{\left (M +1\right )^{L +2}} &  =\frac{\lambda M}{\left (M +1\right )^{2}} \\
\lambda  &  =\frac{\left (M +1\right )^{2}}{M}\end{align}
which leads to
\begin{equation}p_{i}\left (L\right ) =\frac{1}{M\left (M +1\right )^{L}}\qquad i =1 ,\ldots  ,M^{L}
\end{equation}
Now, defining the rank of a given word
$v_{k}(L)$
as
$r(L) ,$
then after performing an exponential transformation from the word length to word rank,
the
probability of occurrence of a given word in terms of rank can be defined as \cite{Montemurro2001},\cite{Mandelbrot83}:
\begin{equation}p(r) =\frac{\gamma _{}}{(r +\beta )^{\alpha }} \label{prank}
\end{equation}
where, for iid samples,
Li showed the constants can be computed as  \cite{li92random}:
\begin{equation}\alpha  =\frac{\log _{2}(M +1)}{\log _{2}(M)} ,\beta  =\frac{M}{M +1} ,\gamma _{} =\frac{M^{\alpha  -1}}{(M -1)^{\alpha }}
\end{equation}
We introduce a normalization step as follows. Since
\begin{equation}\sum \limits _{i =1}^{M}p(i) =1 ,\quad \sum \limits _{i =1}^{\infty }\frac{\gamma _{}}{(r +\beta )^{\alpha }} =\kappa 
\end{equation}
we introduce
\begin{equation}\gamma ^{ \prime } =\frac{\gamma }{\kappa }
\end{equation}
which leads to
\begin{equation}p(r) =\frac{\gamma ^{ \prime }}{(r +\beta )^{\alpha }} \label{prank2}
\end{equation}
Now we have
\begin{equation}p(r) =\frac{M^{\alpha  -1}_{}}{\left [\left (M -1\right )(r +\beta )^{}\right ]^{\alpha }}
\end{equation}
We seek to determine
\begin{equation}\varphi _{0}(x ,r) =\inf _{r}\left \{\left \vert p(x ,r -1) -p(x ,r)\right \vert  ,x \in X ,r \in [1 ,M]\right \} \label{p0inf}
\end{equation}
Hence we define,
\begin{equation}\theta _{j} =d_{j} -d_{j +1} ,
\end{equation}
where
\begin{align}d_{j} &  =\frac{p(r -j) -p(r -j +1)}{\gamma ^{ \prime ^{}}} \\
 &  =\frac{1}{(r -j +\beta )^{\alpha }} -\frac{1}{(r -j +1 +\beta )^{\alpha }}\end{align}
then for
$M \geq 1 ,$
$\alpha  >$
$1 ,$
and hence
$\widehat{\theta }_{j} >\theta _{j}$
, where we seek to establish whether
$\widehat{\theta }_{j} >0$
(it is not).            For notational convenience, let
$\phi _{j} =r -j +\beta  ,$
then we define
\begin{equation}\widehat{d}_{j} =\frac{1}{\phi _{j}} -\frac{1}{\phi _{j -1}}
\end{equation}
and
\begin{align}\overset{}{_{}}\widehat{\theta }_{j} &  =\left (\frac{1}{\phi _{j}} -\frac{1}{\phi _{j -1}}\right ) -\left (\frac{1}{\phi _{j +1}} -\frac{1}{\phi _{j}}\right ) \\
 &  =\frac{\phi _{j -1} -\phi _{j}}{\phi _{j}\phi _{j -1}} -\frac{\phi _{j} -\phi _{j +1}}{\phi _{j}\phi _{j +1}} \\
 &  =\frac{\phi _{j +1}(\phi _{j -1} -\phi _{j})_{}}{\phi _{j}\phi _{j -1}\phi _{j +1}} -\frac{\phi _{j -1}\left (\phi _{j} -\phi _{j +1}\right )_{}}{\phi _{j}\phi _{j -1}\phi _{j +1}}\end{align}
Hence we define,
\begin{align}\widehat{w}_{j} &  =\phi _{j +1}(\phi _{j -1} -\phi _{j}) -\phi _{j -1}(\phi _{j} -\phi _{j +1}) \\
 &  =(\phi _{j +1} -\phi _{j -1})\end{align}
as
$\phi _{j -1} -\phi _{j} =1 ,$
and
$\phi _{j} -\phi _{j +1} =1.$
Therefore,
\begin{align}\widehat{w}_{j} &  =(r -\left (j +1\right ) +\beta ) -(r -\left (j -1\right ) +\beta ) \\
 &  = -2\end{align}
Hence,
$\widehat{\theta }_{j} <0$
and since
$\widehat{\theta }_{j} >\theta _{j}$
, therefore
$\theta _{j} <0$
. Hence it follows that
$\widehat{d}_{j} <\widehat{d}_{j +1}$,
and
\begin{equation}\frac{p(r -j) - p(r -j +1)}{\gamma ^{ \prime }} <\frac{p(r -\left (j +1\right )) -p(r -j)}{\gamma ^{ \prime }}
\end{equation}
By induction,  we have\begin{equation}\varphi _{0}(x ,r) =p\left (M -1\right ) -p(M) \label{varphi}
\end{equation}
which provides a result which we use in the following to determine an expression
for
$N_{0}$. Hence, using empirical probabilities we can solve (\ref{deltaeqn}) as
\begin{equation}\Delta _{0} =\widehat{p}\left (M -1\right ) -\widehat{p}(M)
\end{equation}
Now, since
\begin{equation}\left .\sup \left \{\widehat{p}(r) +\frac{\Delta _{0}}{2}\right \} =\inf \left \{\widehat{p}(r -1) -\frac{\Delta _{0}}{2}\right \}\right \vert _{r =M} ,
\end{equation}
then we have
\begin{equation}\epsilon _{r}^{} \geq \frac{\Delta _{0}}{4}
\end{equation}
Hence, the value of
$n_{0}$
can be calculated using the DKW inequality and the result in
(\ref{numberN}) as
follows.
\begin{align}n_{0} &  =\frac{1}{2\epsilon _{r}^{2}}\ln \genfrac{(}{)}{}{}{2}{1 -\zeta } \label{n0eqn} \\
 &  \leq \frac{1}{2\left (\frac{\Delta _{0}}{4}\right )^{2}}\ln \genfrac{(}{)}{}{}{2}{1 -\zeta } \\
 &  \leq \frac{8}{\Delta _{0}^{2}}\ln \genfrac{(}{)}{}{}{2}{1 -\zeta } \label{n0eqn3}\end{align}
Hence, from (\ref{mainN}), the minimum number of observations required to estimate the entropy            is
\begin{equation}N_{0} =\frac{n_{0}}{\widehat{p}_{0}(r)} \label{BigNoEqn}
\end{equation}
where,
\begin{align}\widehat{p}_{o}(r) &  =\widehat{p}(M) \\
 &  =\frac{M^{\alpha  -1}_{}}{\left [\left (M -1\right )(M +\beta )^{}\right ]^{\alpha }}\end{align}
Now, this gives the number of observations
required to estimate the entropy within a specified degree of confidence and within
specified bounds by considering the smallest probabilities used. This algorithm is suitable for
estimating the minimum number of samples required to compute the \textit{N}-gram entropy of a sequence.

\subsection{Samples required for coarse entropy classification
}

Suppose we wish to detect major differences
between entropies
$H_{i}$
,
$H_{i +1}$
due to changes in the \textit{most
frequent} symbol probabilities. What then is the minimum number of samples
$N_{0}$
required? In this case, to find
$N_{0}$
for
$\sup (\epsilon _{0}),$
implies detecting changes due to
$p(r),p(r +1),\ldots $
where
$r \ll M$
, e.g.
$r =1 ,2.$
Hence define a new alphabet size
$M_{c} <M,$
then since
\begin{equation}\sum \limits _{i =1}^{M_{c}}\widehat{p}(i) \approx \sum \limits _{i =1}^{M_{}}\widehat{p}(i)
\end{equation}
we may omit
$\left \{\widehat{p}(i)\right \}$
$ \forall i >M_{c}$
(i.e. the most infrequently occurring symbols), and thus we select
\begin{equation}\Delta _{0} =\widehat{p}\left (M_{c} -1\right ) -\widehat{p}(M_{c})
\end{equation}
where the final number of samples may be approximated directly from (\ref{n0eqn3}) and (\ref{BigNoEqn})
substituting
$M_{c}$
for
$M$
. Similarly, this admits other related approaches to varying
$\Delta _{0}$
for example, the top
$q \%$
of the probabilities, e.g.
\begin{equation}r =\left \lceil \frac{M}{4}\right \rceil 
\end{equation}

An example of using this approach is given in the next section.

\section{Example results
}

\subsubsection{\textit{First order entropy of
}\textit{English text         }}
 Consider the following example of calculating the number of samples required to
determine the entropy\protect\footnote{ It has been debated in the literature as to how closely English
letters follow Zipf's law and in addition, noted that such measurements may be subject to
observational bias \cite{liw02},\cite{Corral2015}.
}. Let the alphabet size be
$M =26 ,$
and confidence level of the calculation be 95\%, i.e.
$\zeta ^{ \prime } =0.95.$
Hence,
\begin{equation}n =\frac{8}{\Delta _{0}^{2}}\ln \genfrac{(}{)}{}{}{2}{1 -0.95}
\end{equation}
where the probabilities are computed from (\ref{prank2}) as
\begin{equation}\widehat{p}(26) =\frac{\gamma _{}^{ \prime ^{}}}{(26 +\beta )^{\alpha }}
\end{equation}
with the parameters computed as:
\begin{align}\alpha  =\frac{\log _{2}(26 +1)}{\log _{2}(26)} \approx 1.012 ,\quad \beta  =\frac{26}{26 +1} \approx 0.963 , \\
\gamma ^{ \prime } =\frac{\kappa 26^{\alpha  -1}}{(26 -1)^{\alpha }} \approx 0.351\end{align}
Hence, it can be found that
$\Delta _{0}^{} =4.88 \times 10^{ -4},$
and therefore
$n =1.24 \times 10^{8}$. Now, this implies the total number of observations required is then

\begin{equation}\widehat{N}_{0} =\frac{1.24 \times 10^{8}}{\widehat{p}(26)} =1.06 \times 10^{9}
\end{equation}
Thereby, this gives an estimate of the number of observation samples which may be
required in order to obtain an estimate of the entropy which takes into account the smallest
contributing probabilities, i.e. the most infrequent symbols. The value of approximately 1 billion
samples appears to be consistent with reports in the literature, and provides a useful indication of
the upper bound required to compute entropy in this case.

\subsubsection{\textit{Coarse entropy classification for small alphabet size }}
 Consider the following example of
calculating the number of samples required to detect the difference
between entropies by finding the least number of samples required to detect \textit{major} changes in entropy due to changes in the
\textit{most frequent} symbol probabilities. Let the
alphabet size of interest be
$M =3 ,$
and confidence level of the calculation be 75\%, i.e.
$\zeta ^{ \prime }$
$ =0.75.$
Hence,
\begin{equation}n =\frac{8}{\Delta _{0}^{2}}\ln \genfrac{(}{)}{}{}{2}{1 -0.75}
\end{equation}
where the probabilities are computed from
(\ref{prank}) as before, except that we now
consider the minimal number of samples to detect some of the most frequently observed
symbols, hence we choose
$\Delta _{0} =\widehat{p}\left (1\right ) -\widehat{p}(2)$
, where
\begin{equation}\widehat{p}(2) =\frac{\gamma _{}^{ \prime ^{}}}{(2 +\beta )^{\alpha }}
\end{equation}
with the parameters computed as:
\begin{align}\alpha  =\frac{\log _{2}(3 +1)}{\log _{2}(3)} \approx 1.262 ,\quad \beta  =\frac{3}{3 +1} \approx 0.75 , \\
^{}\gamma ^{ \prime } =\frac{\kappa ^{\alpha  -1}}{(3 -1)^{\alpha }} \approx 1.04\end{align}
It can be found that
$\Delta _{0}^{} =0.223$
and therefore
$n =334.$
Now, this implies the total number of observations required is then
\begin{equation}\widehat{N}_{0} =\frac{334}{\widehat{p}(2)} =1151
\end{equation}
Hence, this gives an estimate of the number of samples required
in order to distinguish two different natural sequences using the most frequently occurring
symbols.

\begin{figure}\centering \includegraphics[bb= 21 205 580 594, width=8cm,]{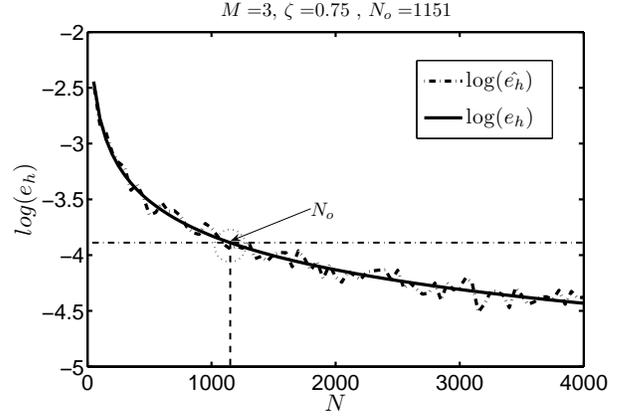}\caption{Performance of the algorithm is shown here using the mse
between the estimated and true entropy as a function of samples
$N$
shown on a log scale. The estimated number of samples required is found at
$N_{0}$
with corresponding mse.
}\end{figure}

This result is demonstrated by simulation in Figs. 1 and 2. In this case,  a small
alphabet size of
$M =3$
was selected which admits a set of probabilities following a Zipfian law. Using ensemble
averaging, a set of data giving rise to empirical probabilities and hence entropies was generated
for a series of differing sample values
$(N) .$
The mean square error (mse) between the true and estimated entropies was obtained as a
function of the number of samples. The efficacy of the method can then be readily observed
particularly in the linear scale plot (Fig  2).
\begin{figure}\centering \includegraphics[bb= 58 205 580 594, width=8cm,]{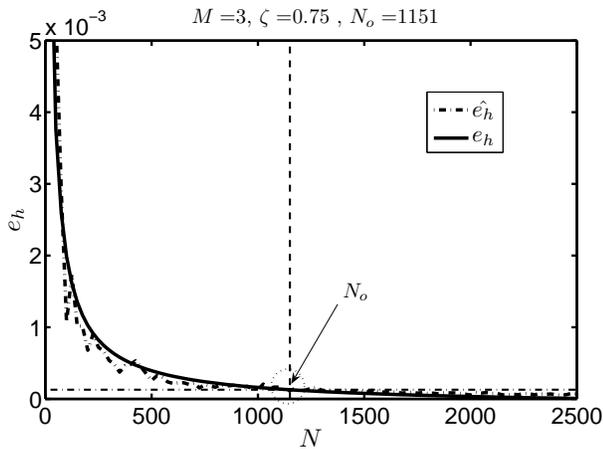}\caption{Performance of the algorithm is shown here using the mse
between the estimated and true entropy as a function of samples
$N$
shown on a linear scale. The estimated number of samples required is found at
$N_{0} ,$
with corresponding mse.
}\end{figure}

\section{Conclusions
}
Shannon entropy is a well known method of measuring the information
content in a sequence of probabilistic symbolic events. In this paper, we have proposed a method of
estimating the number of samples required to estimate the Shannon Entropy for natural sequences.
Using a modified Zipf-Mandelbrot-Li law and the Dvoretzky-Kiefer-Wolfowitz inequality, we propose a
model which yields an estimate for the minimum number of samples required to obtain an estimate
of entropy with a given confidence level and degree of accuracy. Examples have been given which show
the efficacy of the proposed methodology. It would be of interest to apply this
method to various real world applications to compare the theoretical results against experimentally
obtained results. In terms of information theoretic analytical tools, it may be of interest to
consider just how few samples may be required in order to obtain useful results.
Future improvements may possible by re-considering some of the assumptions, such as iid
samples for the parametrization of the Zipf-Mandelbrot-Li model.

\bibliographystyle{IEEEtran}

\begin{thebibliography}{10}
\providecommand{\url}[1]{#1}
\csname url@samestyle\endcsname
\providecommand{\newblock}{\relax}
\providecommand{\bibinfo}[2]{#2}
\providecommand{\BIBentrySTDinterwordspacing}{\spaceskip=0pt\relax}
\providecommand{\BIBentryALTinterwordstretchfactor}{4}
\providecommand{\BIBentryALTinterwordspacing}{\spaceskip=\fontdimen2\font plus
\BIBentryALTinterwordstretchfactor\fontdimen3\font minus
  \fontdimen4\font\relax}
\providecommand{\BIBforeignlanguage}[2]{{%
\expandafter\ifx\csname l@#1\endcsname\relax
\typeout{** WARNING: IEEEtran.bst: No hyphenation pattern has been}%
\typeout{** loaded for the language `#1'. Using the pattern for}%
\typeout{** the default language instead.}%
\else
\language=\csname l@#1\endcsname
\fi
#2}}
\providecommand{\BIBdecl}{\relax}
\BIBdecl

\bibitem{Shannon48}
C.~E. Shannon, ``A mathematical theory of communication (parts {I} and {II}),''
  \emph{Bell System Technical Journal}, vol. XXVII, pp. 379--423, 1948.

\bibitem{Shannon483}
------, ``A mathematical theory of communication (part {III}),'' \emph{Bell
  System Technical Journal}, vol. XXVII, pp. 623--656, 1948.

\bibitem{Shannon51}
------, ``Prediction and entropy of printed {E}nglish,'' \emph{Bell System
  Technical Journal}, pp. 50--64, 1951.

\bibitem{Barnard55}
G.~Barnard, ``Statistical calculation of word entropies for four western
  languages,'' \emph{IRE Transactions Inf. Theory}, pp. 49--53, March 1955.

\bibitem{Herrera2008}
J.~Herrera and P.~Pury, ``Statistical keyword detection in literary corpora,''
  \emph{Eur. Phys. J. B}, vol.~63, pp. 135--146, 05 2008.

\bibitem{Allen2009}
B.~Allen, M.~Kon, and Y.~Bar-Yam, ``A new phylogenetic diversity measure
  generalizing the shannon index and its application to phyllostomid bats,''
  \emph{American Naturalist}, vol. 174, no.~2, pp. 236--243, 2009.

\bibitem{Rao1982}
C.~Rao, ``Diversity and dissimilarity coefficients: a unified approach,''
  \emph{Theoretical Population Biology}, vol.~21, pp. 24--43, 1982.

\bibitem{LeePic2010}
R.~Lee, P.~Jonathan, and P.~Ziman, ``Pictish symbols revealed as a written
  language through application of shannon entropy,'' \emph{Proc. R. Soc. A},
  vol. 466, pp. 2545--2560, 2010.

\bibitem{Liris2011}
R.~A. {Khan}, A.~{Meyer}, H.~{Konik}, and S.~{Bouakaz}, ``{Facial Expression
  Recognition using Entropy and Brightness Features },'' in \emph{11th Int.
  Conf. on Intell. Sys. Design and Applic.(ISDA)}, IEEE, Ed., Dec 2011.

\bibitem{Fuhrman:2000:ASE}
S.~Fuhrman, M.~J. Cunningham, X.~Wen, G.~Zweiger, J.~J. Seilhamer, and
  R.~Somogyi, ``The application of {Shannon} entropy in the identification of
  putative drug targets,'' \emph{Biosystems (A6E)}, vol.~55, no. 1--3, pp.
  5--14, 2000.

\bibitem{Ebeling94}
W.~Ebeling and T.~P\"{o}schel, ``Entropy and long-range correlations in
  literary {E}nglish,'' \emph{Europhysics Letters}, vol.~26, no.~4, p. 241,
  1994.

\bibitem{Moreno2015}
I.~Moreno-S{\'a}nchez, F.~Font-Clos, and {\'A}.~Corral, ``Large-scale analysis
  of {Z}ipf's law in {E}nglish texts,'' \emph{PLOS ONE}, vol.~11, no.~1, 01
  2016.

\bibitem{schurmann-grassberger-96}
T.~Sch{\"u}rmann and P.~Grassberger, ``Entropy estimation of symbol
  sequences,'' \emph{Chaos}, vol. 6(3), pp. 414--427, 1996.

\bibitem{Grassberger1988}
P.~Grassberger, ``Finite sample corrections to entropy and dimension
  estimates,'' \emph{Physics Letters A}, vol. 128, no.~67, pp. 369--373, 1988.

\bibitem{Bonachela2008}
J.~A. Bonachela, H.~Hinrichsen, and M.~A. Mu\~{n}oz, ``Entropy estimates of
  small data sets,'' \emph{Journal of Physics A: Mathematical and Theoretical},
  vol.~41, no. 202001, pp. 1--9, 2008.

\bibitem{Montalvao2012}
J.~Montalv\~{a}o, D.~Silva, and R.~Attux, ``Simple entropy estimator for small
  datasets,'' \emph{Electronics Letters}, vol.~48, pp. 1059--1061, Aug 16 2012.

\bibitem{Paavola2011}
M.~Paavola, ``An efficient entropy estimation approach,'' Ph.D. dissertation,
  University of Oulu, 2011.

\bibitem{lesne2009}
A.~Lesne, J.-L. Blanc, and L.~Pezard, ``Entropy estimation of very short
  symbolic sequences,'' \emph{Phys. Rev. E}, vol.~79, p. 046208, Apr 2009.

\bibitem{dwk56}
A.~Dvoretzky, J.~Kiefer, and J.~Wolfowitz, ``Asymptotic minimax character of
  the sample distribution function and of the classical multinomial
  estimator,'' \emph{Ann. Math. Statist.}, vol.~27, pp. 642--669, 1956.

\bibitem{doob1949}
J.~L. Doob, ``Heuristic approach to the {K}olmogorov-{S}mirnov theorems,''
  \emph{Ann. Math. Statist.}, vol.~20, no.~3, pp. 393--403, 09 1949.

\bibitem{massart1990}
P.~Massart, ``The tight constant in the {D}voretzky-{K}iefer-{W}olfowitz
  {I}nequality,'' \emph{The Annals of Probability}, vol.~18, no.~3, pp.
  1269--1283, 07 1990.

\bibitem{Zielinski2007}
R.~Zielinski, ``Kernel estimators and the {D}voretzky-{K}iefer-{W}olfowitz
  {I}nequality,'' \emph{Applicationes Mathematicae}, vol.~34, pp. 401--404, 01
  2007.

\bibitem{Learned2008}
E.~Learned-Miller and J.~DeStefano, ``A probabilistic upper bound on
  {D}ifferential {E}ntropy,'' \emph{IEEE Trans. on Information Theory},
  vol.~54, pp. 5223 -- 5230, 12 2008.

\bibitem{Zipf1935}
G.~Zipf, \emph{The psycho-biology of language: An introduction to dynamic
  philology}.\hskip 1em plus 0.5em minus 0.4em\relax Cambridge, MA: Houghton
  Mifflin, 1935.

\bibitem{Piantadosi2014}
S.~T. Piantadosi, ``{Z}ipf's word frequency law in natural language: A critical
  review and future directions,'' \emph{Psychonomic Bulletin {\&} Review},
  vol.~21, no.~5, pp. 1112--1130, 2014.

\bibitem{li92random}
W.~Li, ``Random texts exhibit {Z}ipf's-law-like word frequency distribution,''
  \emph{IEEE Transactions on Information Theory}, vol.~38, no.~6, pp.
  1842--1845, 1992.

\bibitem{liw02}
------, ``{Zipf}'s law everywhere,'' \emph{Glottometrics}, vol.~5, pp. 14--21,
  2002.

\bibitem{Montemurro2001}
M.~A. Montemurro, ``Beyond the {Z}ipf-{M}andelbrot law in quantitative
  linguistics,'' \emph{Physica A}, vol. 300, pp. 567--578, Nov 2001.

\bibitem{Corral2015}
{\'A}.~Corral, G.~Boleda, and R.~Ferrer-i Cancho, ``Zipf's law for word
  frequencies: Word forms versus lemmas in long texts,'' \emph{PloS one},
  vol.~10, no.~7, p. e0129031, 2015.

\bibitem{Cancho01}
R.~{Ferrer i Cancho} and R.~V. Sol{\'e}, ``The small-world of human language,''
  \emph{Proceedings of the Royal Society of London B}, vol. 268, no. 1482, pp.
  2261--2265, November 2001.

\bibitem{ChenLeim86}
Y.-S. Chen and F.~Leimkuhler, ``A relationship between {L}otka's law,
  {B}radford's law, and {Z}ipf's law,'' \emph{Journal of the American Society
  for Information Science.}, vol.~37, no.~5, pp. 307--314, 1986.

\bibitem{ChenLeim90}
------, ``Booth's law of word frequency,'' \emph{Journal of the American
  Society for Information Science.}, vol.~41, no.~5, pp. 387--388, 1990.

\bibitem{BOOTH1967}
A.~D. Booth, ``A law of occurrences for words of low frequency,''
  \emph{Information and Control}, vol.~10, no.~4, pp. 386--393, 1967.

\bibitem{Mandelbrot83}
B.~Mandelbrot, \emph{The fractal geometry of nature}.\hskip 1em plus 0.5em
  minus 0.4em\relax New York: W. H. Freeman, 1983.

\end{thebibliography}

\end{document}